\documentclass[%
 aip,
 amsmath,amssymb,
 reprint,
 nofootinbib]{revtex4-2}

\usepackage{graphicx}
\usepackage{dcolumn}
\usepackage{bm}
\usepackage{adjustbox}
\usepackage[utf8]{inputenc}
\usepackage[T1]{fontenc}
\usepackage{mathptmx}
\usepackage{xcolor}
\usepackage{ulem}
\usepackage{hyperref}
\usepackage{tikz}
\usepackage{multirow}

\hypersetup{
  colorlinks   = true,
  citecolor    = blue,
  linkcolor    = blue,
  urlcolor     = blue
}

\newcommand{\mv}[1]{\mathbf{#1}}

\makeatletter
\newcommand*{\citenumns}[2][]{%
  \begingroup
  \let\NAT@mbox=\mbox
  \let\@cite\NAT@citenum
  \let\NAT@space\NAT@spacechar
  \let\NAT@super@kern\relax
  \renewcommand\NAT@open{}%
  \renewcommand\NAT@close{}%
  \cite[#1]{#2}%
  \endgroup
}
\makeatother

\begin{document}

\title{Upscaling DFT-trained machine-learning interatomic potential\\ 
toward Quantum Monte Carlo accuracy:\\
Sulfur-vacancy migration in monolayer MoS\textsubscript{2} as a testbed}

\author{Adam Hlo{\v z}n{\'y}}
\email{adam.hlozny@savba.sk}
\affiliation{Institute of Informatics, Slovak Academy of Sciences, 845~07 Bratislava, Slovakia}
\affiliation{Faculty of Informatics and Information Technologies, Slovak University of Technology, 842~16 Bratislava, Slovakia}

\author{J{\'a}n Brndiar}
\affiliation{Institute of Informatics, Slovak Academy of Sciences, 845~07 Bratislava, Slovakia}

\author{Ye Luo}
\affiliation{Computational Science Division, Argonne National Laboratory, Lemont IL 60439 U.S.A.}

\author{Ivan {\v S}tich}
\email{ivan.stich@savba.sk}
\affiliation{Institute of Informatics, Slovak Academy of Sciences, 845~07 Bratislava, Slovakia}
\affiliation{Institute of Physics, Slovak Academy of Sciences, 84511~Bratislava, Slovakia}

\begin{abstract}

We designed a procedure to train a machine learning interatomic potential (MLIP) at benchmark-quality quantum Monte Carlo (QMC) accuracy. To avoid the complexities of high-quality atomic force determination with the stochastic QMC methods, we use a multi-fidelity approach wherein high-level QMC energies are used alongside suitably processed low-level DFT atomic forces to train a QMC fine-tuned MLIP which significantly improves both the energetics and atomic forces over the baseline DFT-based MLIP.  Fine-tuning is only applied to the readout layers of an equivariant message-passing MACE MLIP. We used sulfur mono- and multiple vacancies in monolayer MoS\textsubscript{2} as a testbed and demonstrate a near QMC accuracy of the model in a number of in- and out-of-domain tests. We show that a fairly limited dataset of QMC energies suffice to significantly improve the baseline DFT MLIP. The accuracy of our approach is demonstrated on energy and free energy migration barriers of mono- and multiple S-vacancy defects. The results open the window to large-scale near QMC quality simulations with large numbers of atoms and/or molecular dynamics configurations which would not be possible by a direct brute-force application of QMC methods.
\end{abstract}

\maketitle

\section{Introduction}
Accurate modeling of potential-energy surfaces (PESs) is of paramount importance as they govern all atomic-scale processes from structural relaxation to molecular dynamics (MD). Activated processes in materials, such as vacancy diffusion, dislocation glide, phase transformations, or chemical reactions at interfaces play a very important role as they sample the most delicate and difficult to describe parts of the PES, the energy or free energy barriers and thus require access to PESs in low-probability regions near transition states.
First-principles methods, such as density functional theory (DFT), can provide forces and energies along reaction pathways, but are typically computationally too expensive for a direct extensive sampling, free-energy calculation, or large supercells, i.e., in situations where the modeling is limited either by system size and/or number of sampling configurations.
\textit{Machine-learning interatomic potentials} (MLIPs) mitigate the cost by learning a surrogate PES from first-principles reference data and thus can make long MD trajectories, large-scale nudged elastic band (NEB) calculations, or free-energy estimation with (near) reference accuracy possible.\cite{Behler2007NNPES,Bartok2010GAP,Behler2016MLIPPerspective}
A critical limitation is that the MLIP accuracy is inherited from the reference method: DFT-trained potentials reproduce DFT.
For many systems, however, the choice of DFT functional may introduce systematic biasses in system energies, such as the energy barrier height or defect energetics, thus motivating the use of more accurate correlated electronic-structure methods.

\textit{Quantum Monte Carlo} (QMC) methods \cite{Foulkes2001QMC} offer an attractive target accuracy as they often provide benchmark-quality energies approaching \textit{chemical accuracy} for both finite and extended systems with a favorable $\mathcal{O}(N^{3-4})$ scaling with system size $N$, compared to correlated quantum chemistry methods, that scale as $\mathcal{O}(N^{6-7})$. Therefore the QMC methods are routinely used as a ``gold standard'' for solids~\cite{Foulkes2001QMC} and 2D materials.\cite{2d_mat_qmc} However, the stochastic nature of QMC methods and the noisy data they produce introduce specifics which need to be taken into account. Current practical limits of their use are $\approx$1000 electrons and $\approx$1000 configurations if used in a MD. An additional obstacle of stochastic QMC methods is that converged atomic forces are significantly harder to obtain than energies, and within fixed-node diffusion Monte Carlo (FN-DMC) the force problem is widely regarded as particularly hard and still under debate, especially for extended systems.\cite{Nakano2024VMCforces,vanRhijn2022EnergyDerivatives,Slootman2024EthanolQMCforces} 
FN-DMC energies can be computed with systematically reducible stochastic error bars and typically small residual biases (time step, population control, fixed-node error), making them suitable for benchmarking energetics.\cite{Foulkes2001QMC}
By contrast, force estimators in FN-DMC must address additional complications, such as Pulay terms, variance issues, or biases related to the fixed-node approximation. Recent work has demonstrated protocols for accurate QMC forces in selected molecular benchmarks\cite{Slootman2024EthanolQMCforces} and for improved estimators of energy derivatives in real-space DMC,\cite{vanRhijn2022EnergyDerivatives} but robust, routine convergence of FN-DMC forces for complex condensed-phase systems is not yet standard and often the significantly less accurate VMC forces are used instead.~\cite{Nakano2024VMCforces} 
These problems could be solved by use of an accurate MLIP which would yield atomic forces by differentiation, but since forces are an integral part of the MLIP design, the lack of high-accuracy atomic forces is quite troubling, as training with force data greatly improves the stability and quality of the MLIP compared to training to energy alone.

Currently, the most widely used approaches for force inference are based on $\Delta$-learning, where a baseline low-level PES/force field is retained and a second model learns a high-level correction. In molecular benchmarks, this strategy has been used to elevate DFT-based potentials toward CCSD(T) quality, including a correction of the classical force field for ethanol~\cite{Nandi2024DeltaML}. For condensed phases, a short-range $\Delta$-ML was used to correct periodic low-level MLIPs using high-level cluster corrections. This model was further developed into a practical tool for routine CCSD(T)-level simulations of liquid water, including finite-temperature 
properties~\cite{Meszaros2025srDeltaML,ONeill2025DeltaWater}. In the QMC context, $\Delta$-learning has also been combined with QMC data for high-pressure deuterium Hugoniot modeling \cite{Tenti2024QMCdeltaHugoniot}. Recently,\cite{Radova2025-fine-tuning-mace} fine-tuning (FT) with freezing layers has been used on the MACE neural network (NN)~\cite{Batatia2022MACE} to provide improvement of foundational models for large scale calculations. The authors use a layer freezing scheme that keeps the loss function intact and apply the approach to quantum chemistry molecular benchmark datasets.  

These complications call for machine learning methods relying solely on the FN-DMC energy landscape to infer FN-DMC-quality forces. Recently the idea of multi-fidelity learning (MFL) to train a MLIP was proposed and explicitly demonstrated that MFL with low-level forces and high-level energies yields MLIP far more accurate than a single-fidelity (SF) MLIP trained solely to high-level energies and almost as accurate as a SF MLIP trained directly to high-level energies and forces.\cite{messerly_25} We follow that idea and show that MFL \textit{MLIP based on FN-DMC energies} and suitably adapted \textit{DFT forces} in combination with FT indeed lead to design of a very accurate MLIP at a computationally bearable cost. 

Here we constructed a partially frozen NN fine-tuning-based approach. In the spirit of MFL, the method we developed targets a (near) QMC energy and force accuracy by correcting the DFT-trained MLIP using solely the QMC energies and suitably processed DFT forces, while preventing the corrected model from developing unphysical or qualitatively incorrect forces. Sulfur-vacancy migration in monolayer MoS\textsubscript{2} is used as a testbed,\cite{Komsa2015NativeDefectsMoS2,Zhou2013IntrinsicDefectsMoS2} primarily because it provides an archetypal activated process with well-defined near-barrier configurations and because we have previously designed a DFT-based sampling, training pipeline, and a reliable DFT-MLIP for this system.\cite{Hlozny2025MoS2Vacancies}

Design of the FT MLIP involves: a) Generation of an affordable but noisy set of FN-DMC energies on configurations drawn from a targeted DFT-based rare-event sampling protocol, b) Fine-tuning of only the readout layers of an equivariant message-passing MLIP (FT-MLIP), MACE~\cite{Batatia2022MACE} in our case, to match the QMC energies, and, c) Enforcing a constraint that keeps the predicted fine-tuned FN-DMC forces at a controlled distance from the DFT baseline data. The distance is controlled by a threshold parameter, which is set according to physical scales of the problem which, in principle, could be further adjusted algorithmically. We have extensively tested the In-Domain (ID) and Out-Of-Domain (OOD) performance of the FT-MLIP by validating the data by comparison with data explicitly calculated by FN-QMC.

Using the FT-MLIP we have performed a range of large-scale simulations of multiple sulfur defects in monolayer MoS$_{2}$, including up to quad S-defects and thermodynamic integration~\cite{Frenkel_07} to determine the free energy barriers. Such calculations would not be feasible by a brute-force FN-QMC calculation because of the system size and/or the number of configurations which need be sampled.  

\section{Methods}
\subsection{DFT and QMC calculations}
We build on the MACE family of equivariant message-passing NN for interatomic potentials.\cite{Batatia2022MACE}
Given an $N$-atom configuration \(\{\mv r_i, z_i\}_{i=1}^N\), with \(\mv r_i \in \mathbb{R}^{3}\) being the position of atom \(i\), 
\(z_i\) the chemical element, MACE constructs per-atom learnable features \(h_i^{(t)}\) through a sequence of equivariant message passing layers. A \textit{readout} is the final mapping from learned, typically invariant, features to the target scalar(s), here the per-atom energy contributions. In the message passing NN MACE formalism, the total energy is built as a sum of per-atom energies obtained from readout functions applied to the node states across the layers.\cite{Batatia2022MACE}
Forces exerted on atoms are then computed as analytical derivatives of the predicted energy landscape with respect to atomic positions.

The practical implication is that \textit{freezing message-passing layers} preserves the learned geometric representation of local environments and thereby also much of the DFT-learned force field, while adjusting only the readout layers primarily changes the mapping from those representations to energies.
This is the key to our \textit{``energy correction under force control''} strategy.

On our testbed, the sulfur-vacancy (S-vacancy) migration in monolayer 2H-MoS\textsubscript{2}.\cite{Komsa2015NativeDefectsMoS2,Zhou2013IntrinsicDefectsMoS2}
We start from a DFT-trained MACE potential developed in our previous work.\cite{Hlozny2025MoS2Vacancies}
That work introduced a multi-step dataset construction strategy for activated processes consisting of (i) generating a minimum-energy path by Nudged Elastic Band method (NEB),\cite{Henkelman2000NEB} (ii) perturbing the path by a Gaussian-process perturbation to diversify the configurations near the saddle point, and (iii) running constrained MD along a reaction coordinate to densely sample the near-barrier structures.\cite{Carter1989BlueMoon}
The resulting dataset is balanced to cover both in-domain (ID) interpolation and out-of-domain (OOD) near-transition-state configurations. The potential was validated using specialized force metrics focusing on the atoms most relevant to migration.\cite{Hlozny2025MoS2Vacancies} The DFT data was generated using the \texttt{GPAW} code.~\cite{gpaw} All technical details are summarized in Ref. [\citenumns{Hlozny2025MoS2Vacancies}].

In the present work, this DFT-trained potential serves as, (a) a stable baseline for geometry and sampling, (b) a force reference to constrain the FN-DMC-based fine-tuning, and (c) a dataset of configurations, for subset of which the FN-DMC energies are calculated.

\begin{figure}
\centering
\includegraphics[width=0.92\columnwidth]{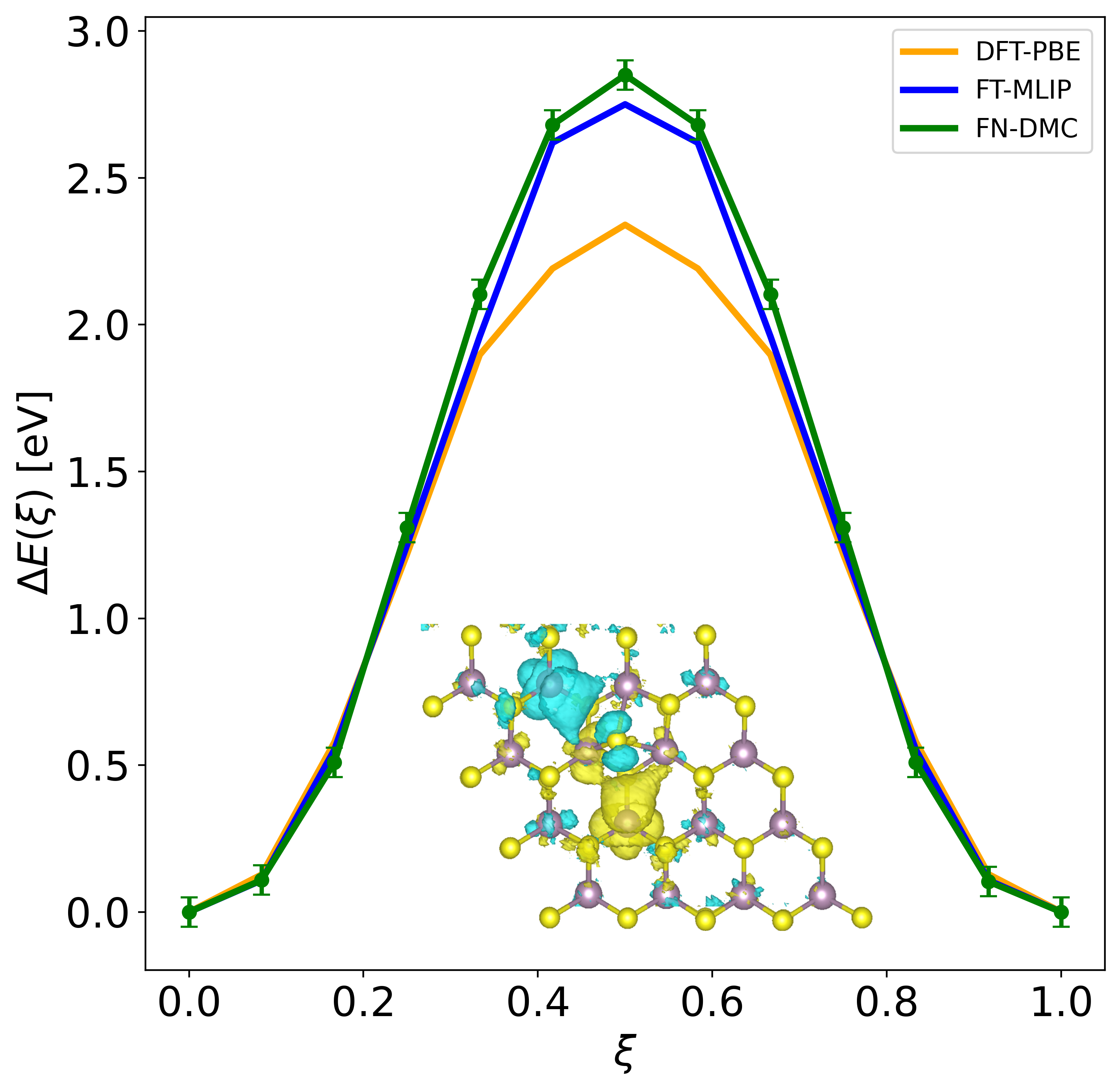}
\caption{\textbf{Calculated migration barrier of S-mono-vacancy in monolayer MoS$_{2}$ in various treatments.} Energy profiles in singlet DFT-PBE, FT-MLIP, and explicit FN-DMC modeling for a single S-vacancy migration in MoS$_{2}$ calculated 
over a DFT-PBE NEB pathway. Note that energy profiles of the antiferromagnetic and singlet DFT-PBE solutions are identical.~\cite{Hlozny2025MoS2Vacancies} The inset shows spin densities near the top of the barrier as sampled by FN-DMC with the antiferromagnetically magnetized Mo atoms and the much smaller spin densities on the sulfur atoms straddling the molybdenum atoms.
\label{fig:QMCDFT_NEB}
}
\end{figure}

In order to highlight the improvement FN-DMC~\cite{Foulkes2001QMC} brings into the defect migration energetics, we show in Fig.~\ref{fig:QMCDFT_NEB} comparison of DFT-PBE and FN-DMC energy barriers for a single S-vacancy calculated perturbatively along the DFT-PBE NEB pathway. FN-DMC calculations were performed with the \texttt{QMCPACK} suite of codes~\cite{qmcpack} in fixed-node approximation, using Slater-Jastrow type of variational Monte Carlo (VMC) trial wave functions. The nodal hypersurfaces were determined by DFT orbitals using the generalized gradient, DFT-PBE, approximation~\cite{pbe}, calculated with the \texttt{Quantum Espresso} codes~\cite{q-espresso}. Short-range correlations were described by Jastrow factor~\cite{Foulkes2001QMC}. Atomic cores were replaced by Effective Core Potentials~\cite{ECPs}. All technical details are summarized in Appendix~\ref{app:qmc-details}. As shown in Fig.~\ref{fig:QMCDFT_NEB}, the quantitative effect of use of the many-body FN-DMC method is increase in the calculated defect migration energy by 0.55 eV compared to DFT-PBE.

To generate the QMC dataset at a manageable cost, we computed FN-DMC single-point energies for $\mathcal{O}(10^3)$ configurations  sampled from the constrained, finite-temperature DFT-PBE MD generated as described above, i.e. for near-minima and near-saddle point configurations for \textit{mono S-vacancy} migration, with 50\% configurations sampled at 400 K and 50\% at 800 K. The configurations are not uniformly optimized at the QMC level. Rather, we intentionally accept moderately converged DMC energies to maximize configuration-space coverage under a fixed computational budget. Details of the FN-DMC data acquisition are summarized in Appendix~\ref{app:qmc-details}.

\subsection{Upscaling the MLIP toward FN-DMC accuracy:\\ 
High-level energies and low-level forces}
\label{sec:upscaling-mlip-toward-qmc-intro}
Instead of calculating of high-accuracy FN-DMC forces~\cite{Nakano2024VMCforces} we use an alternative route for FT-MLIP based on high-level energies and low-level forces. Such objective can be achieved if: a) the FN-DMC and DFT PESs are qualitatively similar and the differences are only quantitative, such as different energy barrier heights, b) the FN-DMC PES is smooth and atomic forces can be calculated via differentiation of the energy, an assumption automatically met by NN-based MLIPs, c) consistency of forces and energies is maintained, and d) a reliable upscaling technique is introduced, where fine tuning MLIPs lend themselves naturally. a) and b) mean that the procedure will only work subject to the condition that the correlation effects \textit{do not} affect the PES qualitatively, meaning also that the procedure will not be entirely general. We now describe c), and the design of the loss function $L$.

Let \(E_{\mathrm{pred}}\) and \(\mv F_{\mathrm{pred}}\) denote energy and forces predicted by the MLIP, \(E_{\mathrm{QMC}}\) the FN-DMC energy, and \(\mv F_{\mathrm{DFT}}\) the DFT forces for the same configuration.
We minimize a combined loss
\begin{equation}
\label{eq:loss}
\begin{aligned}
L(E_{\mathrm{pred}}, \mv F_{\mathrm{pred}};\,
  E_{\mathrm{QMC}}, \mv F_{\mathrm{DFT}})
= &\lambda_E\,\mathrm{MSE}(E_{\mathrm{pred}}, E_{\mathrm{QMC}}) \\
+ &\lambda_F\,\mathrm{FE}_t(\mv F_{\mathrm{pred}}, \mv F_{\mathrm{DFT}})~~~,
\end{aligned}
\end{equation}
where MSE stands for mean squared error and \(\lambda_E\) and \(\lambda_F\) set the energy--force tradeoff.\\
The force term is a thresholded penalty designed to allow the model to deviate from the DFT force field only when necessary and only up to a controlled magnitude.
We define
\begin{equation}
\label{eq:forceerror}
\mathrm{FE}_t(\mv F_{\mathrm{pred}},\mv F_{\mathrm{DFT}})
= \Theta\!\left( \|\Delta \mv F\|^2 - t^2 \right)\,\left( \|\Delta \mv F\|^2 - t^2 \right)^{2}~~~,
\end{equation}
with \(\Delta\mv F = \mv F_{\mathrm{pred}}-\mv F_{\mathrm{DFT}}\), \(\Theta\) the Heaviside step function, and
\begin{equation}
\label{eq:norm}
\|\Delta \mv F\|^2 = \frac{1}{N}\sum_{i=1}^{N}\left\|\mv F^{\mathrm{pred}}_i-\mv F^{\mathrm{DFT}}_i\right\|^2~~~.
\end{equation}
\(t\) in Eq. (\ref{eq:forceerror}) has units of force and can be interpreted as an allowed per atom RMS deviation.

Minimizing Eq.~\eqref{eq:loss} fits the FN-DMC energies while preventing large deviations from the DFT forces.
The threshold \(t\) is the main control parameter; increasing \(t\) relaxes the force constraint, enabling larger QMC-driven corrections but increasing the risk of unphysical forces and unstable MD, while decreasing \(t\) enforces closer adherence to the DFT dynamics with more limited learning of the QMC energy corrections.
In practice, \(t\) should be selected as large as possible while maintaining stable MD, energy improvement on held-out QMC energies and generalization performance. Our empirical findings show, that introduction of \(t\) in Eq. (\ref{eq:forceerror}) is beneficial in preventing overfitting and keeping out-of-domain MD runs stable, but \(t\) should not directly be used to "tune the precision".

d) is achieved by FT a DFT-pretrained MACE foundation model by freezing the equivariant message-passing stack and updating only the energy readout mapping from per-atom invariant features to per-atom energies summed to the total energy.\cite{Batatia2022MACE} This constrained update is used to preserve the priors learned at the DFT level while allowing recalibration to the QMC-informed targets. The model uses \(r_{\max}=4.0~\text{\AA}\), hidden irreps \(128\times 0e + 128\times 1o\), and single-precision training (\texttt{float32}) on GPU.

Training uses AdamW (\(\beta\)-variant with AMSGrad enabled), initial learning rate of \(10^{-3}\), batch size 32, gradient clipping at 10.0, and up to 100 epochs. We apply a \texttt{ReduceLROnPlateau} scheduler with patience of 5 epochs. No stochastic weight averaging, exponential moving average, or weight decay is used in the runs. Atomic reference energies are handled with \(E_0\) set to dataset average.

The variables in the loss function in Eq. (\ref{eq:loss}) were set to \(\lambda_E = 100\), \(\lambda_F = 5\) and the force threshold in Eq. (\ref{eq:forceerror}) to \(t = 16\) eV/\AA. The FT-MLIP is available for download.~\cite{FT-MLIP_github}

\section{Results and Discussion}

\textit{QMC-force benchmarking.} To assess whether the force-constrained FT improves forces relative to QMC without relying on direct QMC force estimators, we construct a limited benchmark of force components using the FN-DMC energies only.
For three atoms most critical for the vacancy migration selected from the set used for the RMSE\textsubscript{5} force metric in Ref.~[\citenumns{Hlozny2025MoS2Vacancies}] we displaced a chosen cartesian coordinate by \(x\in\{-2\Delta,-\Delta,0,\Delta,2\Delta\}\) with \(\Delta=0.08\)~\AA, and computed the FN-DMC energies \(E(x)\).
For each displacement series, we fitted \(E(x)\) by a quadratic model:
\begin{equation}
E(x) = a_0 + a_1 x + a_2 x^2,
\end{equation}
and estimated the force component as
\begin{equation}
F_x^{\mathrm{QMC}} \approx -\left.\frac{dE}{dx}\right|_{x=0} = -a_1~~~,
\label{eq:quad_force}
\end{equation}
with  \(y\) and \(z\) components handled identically.

We performed this procedure for 27 displacement profiles in total: 3 atoms, 3 coordinates, 3 sections of the MD trajectory- near minimum, near transition point, and near the steepest ascent/descent, for details, see Appendix~\ref{app:force-benchmarks}. We then compared these QMC-derived force components to the corresponding components predicted by the baseline DFT-MLIP and the FT-MLIP models. An example for one force component is shown in Fig.~\ref{fig:forcebench}, for the rest, see Appendix~\ref{app:force-benchmarks}. One can see that at all displacements the DFT and FN-QMC energies differ by as much as $\approx$0.3 eV, meaning that the corrugation of the PES is significantly larger in FN-DMC than in the DFT. Such energy differences account also for the difference in the migration barrier heights. On the other hand, the FT-MLIP and FN-DMC energy profiles and all DFT-PBE, FT-MLIP and FN-DMC forces differ only slightly: 2.31 eV/\AA~(FN-DMC) vs. 2.26~eV/\AA~(FT-MLIP) vs. 2.03 eV/\AA~(DFT-MLIP). The force-consistency with energy is mediated by a number of quite small force differences.
For over 27 extracted force components, for comparison see Fig.~\ref{fig:forcebench}, the baseline DFT MLIP yields a mean absolute error (MAE) of \(220\) meV/\AA\ relative to the QMC derivatives, while the FT-MLIP model reduces the MAE to \(160\) meV/\AA.~
These results indicate that, even though the model is not trained on QMC forces, the energy-driven correction combined with force anchoring can provide a measurable improvement in force fidelity as well by primarily improving the largest force components.

\begin{figure}
\centering
\includegraphics[width=0.8\columnwidth]{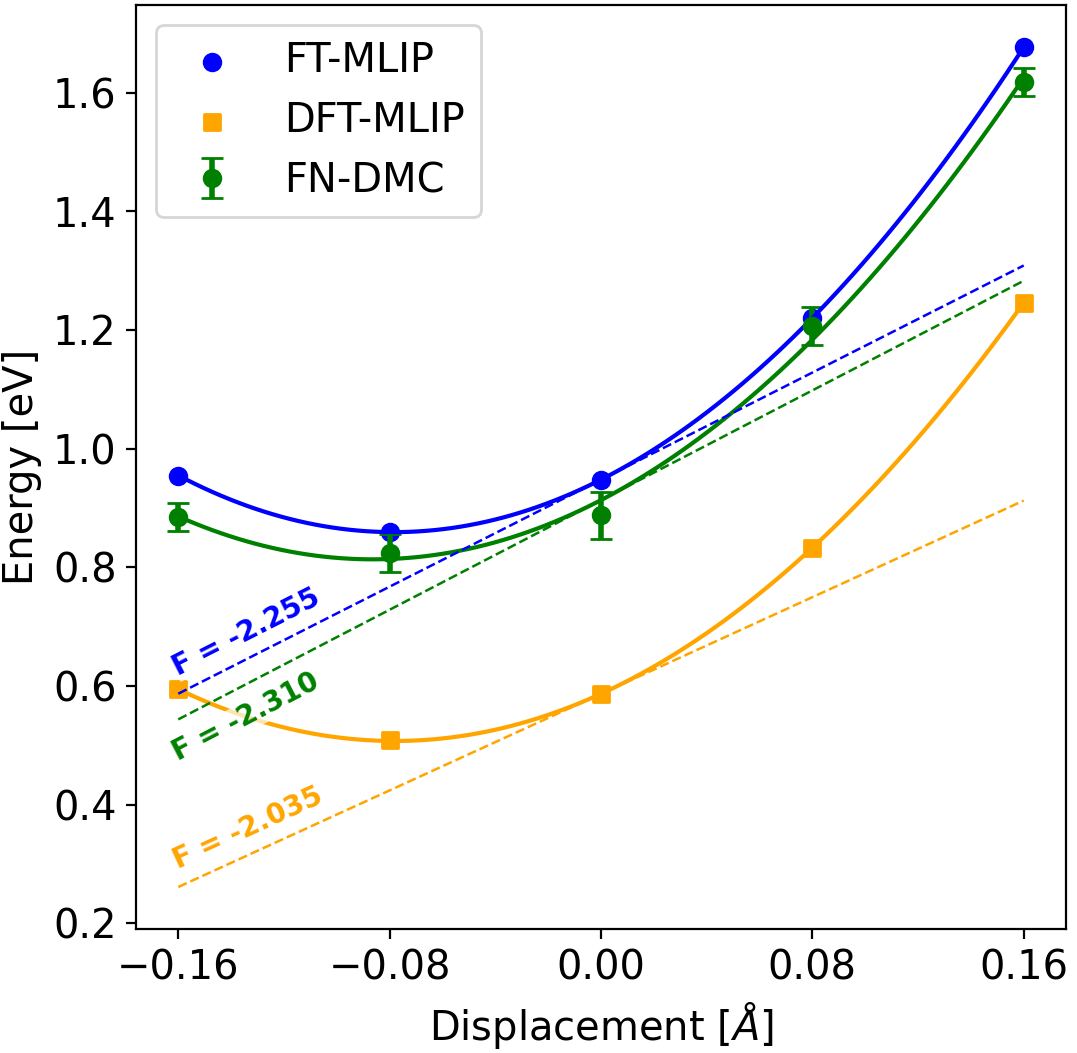}
\includegraphics[width=1.\columnwidth]{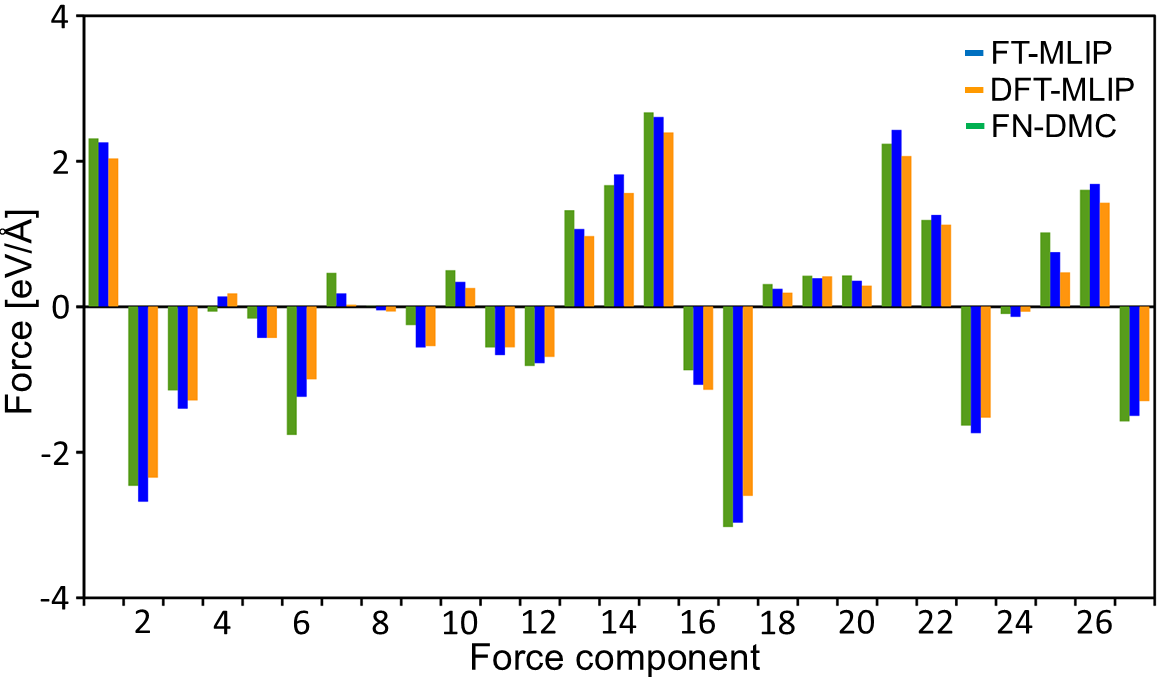}
\caption{\textbf{Force benchmark against FN-DMC energy derivatives.} 
Upper panel: Example of finite-difference benchmark of FN-DMC, FT-MLIP, and DFT-MLIP for one force component. Note that all three energy profiles are energy-aligned at one configuration. 
Lower panel: Comparison of all 27 force components computed via FN-DMC quadratic fit. FT-MLIP, and DFT-MLIP.
\label{fig:forcebench}}
\end{figure}

\begin{figure}
\centering
\includegraphics[width=0.9\linewidth]{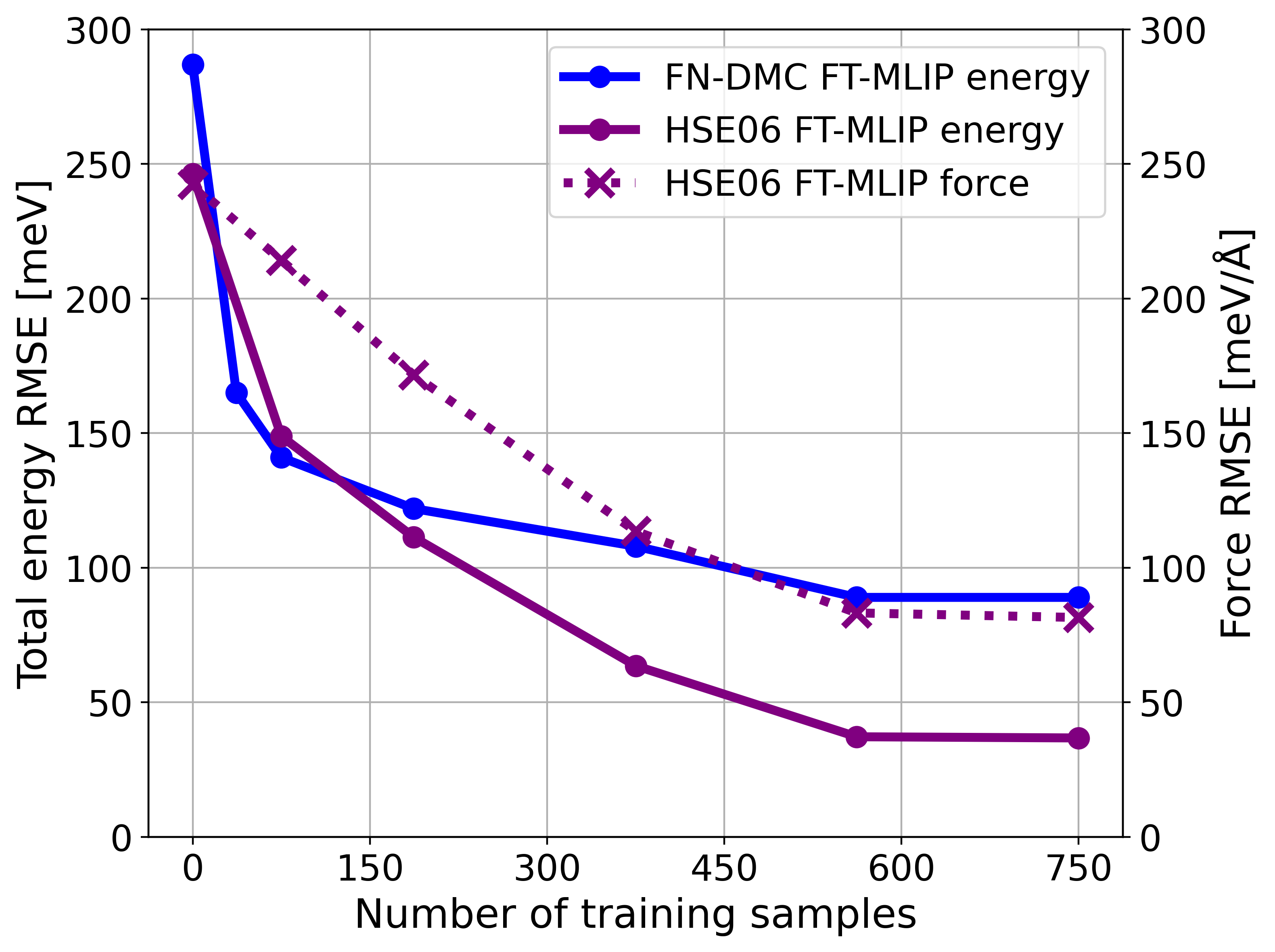}
\caption{\textbf{Data ablation study.} Performance of the model on the validation dataset dependent on the training dataset size. 0 samples means the performance of the baseline model with no fine-tuning. In addition to FN-DMC study we also include a study on data generated using DFT-HSE06 exchange correlation functional as a substitute for FN-DMC in order to show that our procedure produces improvement on forces despite forces not being included in the training dataset.
\label{fig:dataset-ablations}}
\end{figure}

\textit{Dataset ablation study.} Our tuning procedure was based on an arbitrarily chosen size of the QMC tuning dataset dictated primarily by the huge QMC computer budget needed. To test the performance of the approach dependent on the dataset size, we re-trained the model with the same parameters on datasets of different sizes. Each dataset is subset of the full dataset, with the smallest one as small as 37 samples. The results are shown in Fig.~\ref{fig:dataset-ablations}. Most of the accuracy improvement occurs by adding the first 37 QMC-sampled energies to the tuning dataset which reduce the total energy RMSE from $\approx$290 meV to $\approx$160 meV and very little improvement happens with data sizes larger than $\approx$500, where the total energy RMSE stabilizes at 89 meV. To assess directly the performance of the MLIP on atomic forces, we performed an additional study using the hybrid HSE06 exchange correlation functional~\cite{heyd_03,heyd_06} and generated an additional DFT-HSE06 dataset as a proxy for the QMC dataset. During training, HSE06 force information was deliberately excluded and only DFT-HSE06 energies and DFT-PBE atomic forces were used for tuning in a way identical to the MFL tuning we used in the FN-DMC FT. The performance of this HSE06 FT-MLIP on reproduction of atomic forces was then evaluated on the test set. These results are also shown in Fig.~\ref{fig:dataset-ablations}. In general, the dependence on the training dataset size is similar to that for the FN-DMC dataset with the energy RMSE stabilizing at $\approx$35 meV, about half the value for the FN-DMC dataset, presumably due to absence of the noise in the HSE06 data. Despite DFT-HSE06 forces not being included in the training, the resulting model exhibits a significant improvement in force predictions, with atomic force RMSE of $\approx$80 meV/\AA. We have performed a third test by using a SF tuning to only DFT-HSE06 energies. Perhaps surprisingly, the overall energy and force RMSE did not change appreciably. Instead, the improved force prediction quality manifested itself by greatly improved MD stability, preventing crashes due to the occasional inconsistencies between energies and forces. This behavior is most likely caused by freezing most of the NN during FT. The results indicate that the MFL tuning with high-level energies and low-level atomic forces not only improves the energy and force RMSE but, most importantly, it greatly improves the stability of MD runs.

\textit{In-domain tests.}
The only ID test performed was determination of the NEB migration barrier which used the FN-DMC dataset generated for the mono-vacancy. The result is shown in Fig.~\ref{fig:QMCDFT_NEB} and Tab.~\ref{tab:barriers}. The explicitly FN-DMC- and FT-MLIP-calculated barrier heights differ by mere $\approx$0.1 eV. Contrary, the DFT-MLIP barrier is 0.55 eV lower than the explicitly determined FN-DMC barrier and 0.45 eV lower than the FT-MLIP barrier, thus proving the high quality of the MFL high-level energy low-level force FT-MLIP to describe energies.

\begin{figure}
\centering
\includegraphics[width=0.92\columnwidth]{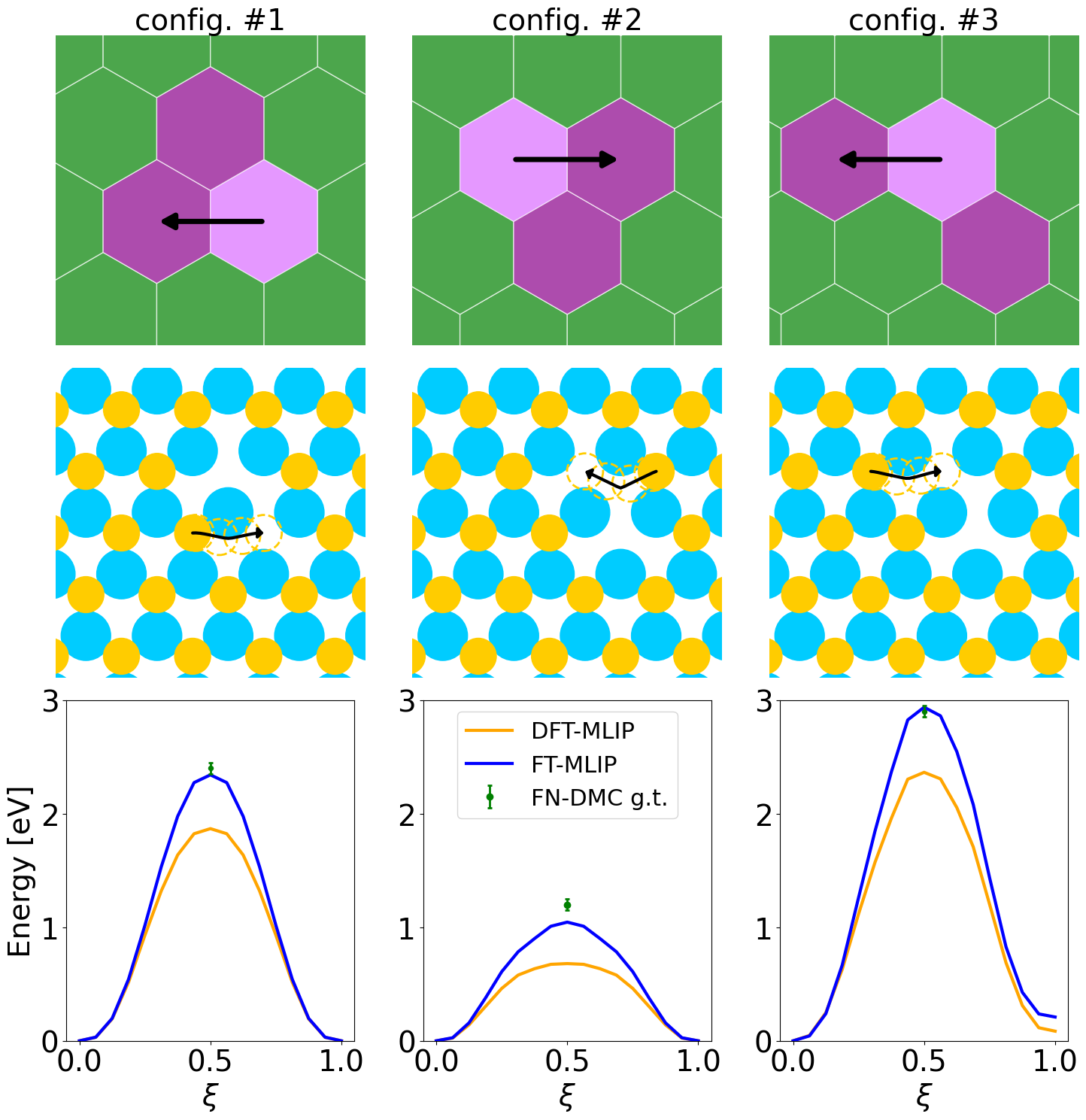}
\caption{\textbf{Migration barriers for bi-vacancies.} Calculation of bi-vacancy barriers for OOD testing. Selected configurations
are shown with each column representing a different configuration. Upper panel: Schematics of the various double defect configurations
together with initial and final position of the vacancy. Middle panel: Visualization of the transition from the view point of the supercell. Lower panel: FT-MLIP energy profiles for given pathways compared to DFT-MLIP baseline model. The cross indicates the explicitly FN-QMC computed point at the transition state. Each calculation is done on 5 $\times$ 5 supercell.
}
\label{fig:bi-vacancy-ood}
\end{figure}

\begin{figure}
\centering
\includegraphics[width=0.92\columnwidth]{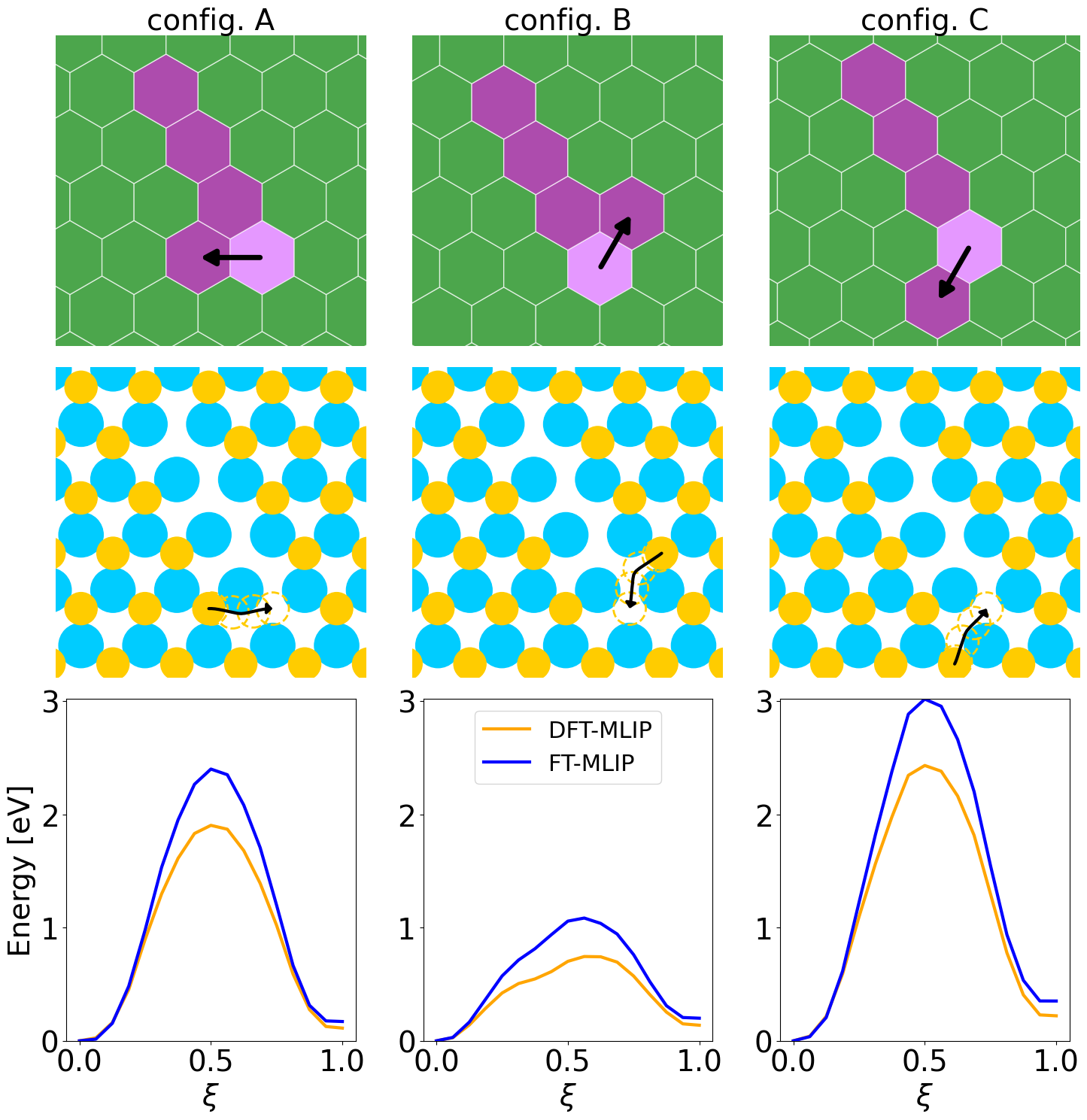}
\caption{\textbf{Migration barriers for quad-vacancies.}  Calculation of quad-vacancy barriers forming a line defect transfer testing. 
Selected configurations are shown with each column representing a different configuration. Upper panel: Schematics of the various double defect configurations together with initial and final position of the vacancy. Middle panel: Visualization of the transition from the view point of the supercell. Lower panel: FT-MLIP energy profiles for given pathways compared to DFT-MLIP baseline model. Each calculation is done on 9 $\times$ 9 supercell. This serves as a transfer test illustrating how QMC-informed corrections may alter rare-event energetics beyond the immediate fine-tuning set.
\label{fig:linedefect}}
\end{figure}

\begin{table}[b!]
\centering
\caption{\textbf{In-, Out-of-domain, and transfer tests for S-vacancy and multi-vacancies in monolayer MoS$_{2}$.} Comparison of energy/free energy barriers for mono-, bi-, and quad- S-vacancies in various configurations, for labeling see Figs.~\ref{fig:bi-vacancy-ood},~\ref{fig:linedefect}, as calculated by DFT-MLIP, FT-MLIP, and explicitly determined by FN-DMC. For each type of system the first line always corresponds to energies, the second to free energies, The three free energy entries correspond to 300, 600, and 900 K.
}
\label{tab:barriers}
\begin{tabular}{l|c|c|c|c}
\hline
\hline
\multirow{2}{*}{system}      &  \multirow{2}{*}{config.} & \multicolumn{3}{c}{$E_{B}$/$F_{B}$ [eV]} \\ \cline{3-5}
                             &      & DFT-MLIP & FN-DMC & FT-MLIP \\
\hline
\hline
\multirow{2}{*}{mono-vacancy}&\multirow{2}{*}{n/a} & 2.30 & 2.85 & 2.75                \\
                             &                     & 2.1~2.0~1.9 & n/a & 2.6~2.5~2.4\\
\hline
\multirow{6}{*}{bi-vacancy}  & \multirow{2}{*}{1}  & 1.86 & 2.38 & 2.34         \\ 
                             &                     & n/a & n/a & 2.1~2.0~1.9 \\ \cline{2-5}
                             & \multirow{2}{*}{2}  & 0.68 & 1.20 & 1.05         \\
                             &                     & n/a & n/a & 0.9~0.8~0.7 \\ \cline{2-5}
                             & \multirow{2}{*}{3}  & 2.36 & 2.89 & 2.93         \\ 
                             &                     & n/a & n/a & 2.8~2.7~2.6 \\
\hline                             
\multirow{3}{*}{quad-vacancy}&                 A   & 1.90 & n/a & 2.40   \\ 
                             &                 B   & 0.74 & n/a & 1.08   \\
                             &                 C   & 2.43 & n/a & 3.02   \\
\hline
\hline
\end{tabular}
\end{table}

\textit{Out-of-domain tests.}
Since our FN-QMC FT dataset comprised only of mono-vacancy, as an OOD test we calculated the FT-MLIP migration barriers for representative bi-vacancy barriers, see Fig. \ref{fig:bi-vacancy-ood} and Tab.~\ref{tab:barriers}. To test the performance of the FT-MLIP we have explicitly recomputed the barriers by direct FN-DMC calculations for configurations selected as described in Ref. [\citenumns{Hlozny2025MoS2Vacancies}]. We find that, while all methods yield the correct order of barrier heights, including the ultra-low energy barrier for configuration \# 2~\cite{Hlozny2025MoS2Vacancies}, the FT-MLIP provides a notable increase of all three migration barriers which show only very minor, from 0.04 to at most 0.15 eV, deviations from barriers explicitly determined by FN-DMC. Contrary, the DFT-MLIP barriers are $\approx$0.5 eV lower than the explicitly FN-DMC determined barriers.

\textit{Transfer tests.}
As an additional qualitative transfer test, we apply the FT-MLIP to configurations associated with line-defect formation/migration pathways previously explored with the DFT MLIP.\cite{Hlozny2025MoS2Vacancies} Fig.~\ref{fig:linedefect} and Tab.~\ref{tab:barriers} show representative energy profiles and barriers, respectively, predicted by the DFT-MLIP and FT-MLIP models along the corresponding reaction coordinate. These calculations were performed in a 9 $\times$ 9 supercell where a direct FN-QMC calculation is not feasible as it would require to treat in excess of two thousand of explicitly correlated electrons. The FT-MLIP model produces systematic increases in relative energies compared to the DFT-MLIP, consistent with the expectation that QMC corrections to DFT energetics can depend on local bonding environments.

\begin{figure}
\centering
\includegraphics[width=0.98\columnwidth]{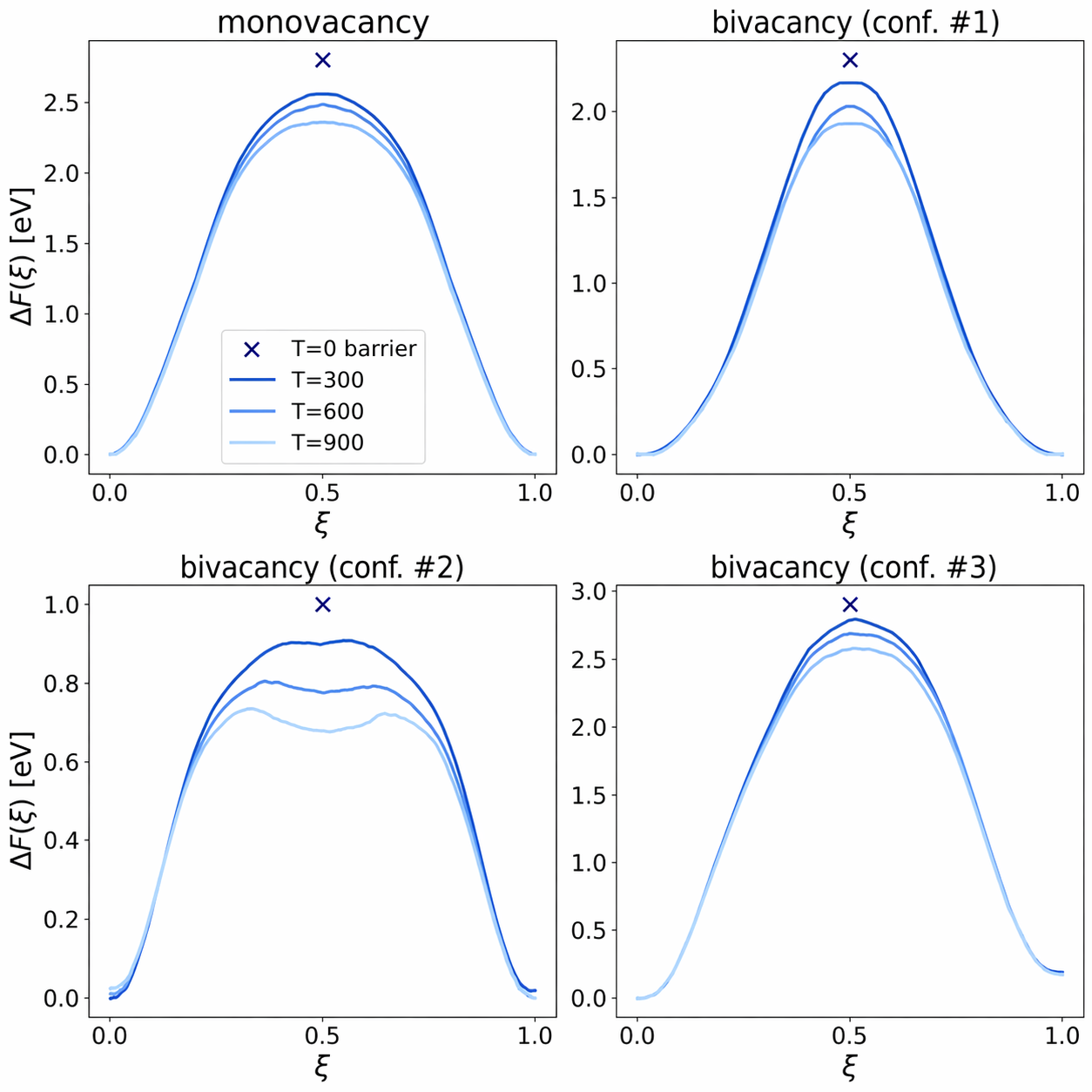}
\caption{\textbf{Simulation of free energies.} Free energy variation along the reaction coordinate $\xi$ for a single and bi- S-vacancy migration at 300, 600, and 900 K. The total energy barrier (T = 0 K) is also shown for comparison. FT-MLIP results were obtained in 
8 $\times$ 8 supercell. 
}
\label{fig:free_mono_and_bi_vac}
\end{figure}

\textit{Finite-temperature MD: Free energies.}
The other demonstration of the power of the FT-MLIP is by showing the ability to perform large-scale MD simulations which require tens of thousands of MD steps in large supercells as is the case with multiple defects. Such simulations conducted by brute-force use of FN-QMC methods would be completely ruled out. We use MD to determine free-energy profiles by methods of thermodynamic integration~\cite{Frenkel_07,Hlozny2025MoS2Vacancies}. We use techniques identical to those used in our previous DFT study~\cite{Hlozny2025MoS2Vacancies} by adiabatically pushing one S-vacancy over the barrier, fixing the position of the vacancy at a given point on the reaction coordinate $\xi$ by the applied constraint. The results for the mono- and bi-vacancy are shown for temperatures of 300, 600, and 900 K in Fig.~\ref{fig:free_mono_and_bi_vac}. The free energy barriers are summarized also in Tab.~\ref{tab:barriers}. At room temperature the vibrational entropy lowers the free energy barrier compared to the total energy barrier by $\approx$0.1--0.2 eV and the free energy barrier lowers by a fixed amount of $\approx$0.1 eV for every temperature increase by 300 K. Comparison with vibrational entropy correction for multi-defects in configurations \# 1 and \# 3 (high energy/free energy barrier) reveals that the vibrational entropy corrections are insensitive to the number of defects and act as a homogeneous add-on correction term. Interestingly, in configuration \# 2 (low energy/free energy barrier), where the defect has more space available due to the strategic location of the other defect~\cite{Hlozny2025MoS2Vacancies}, the free energy at all three temperatures qualitatively modifies the energy profile and a local minimum develops separating two new transition states, both moved toward the ''reactants``/''products``. 
Since the vibrational entropy corrections are comparable to the corrections induced by FT of the DFT results to FN-DMC accuracy, they should be applied simultaneously in any finite-temperature situation.

\section{Conclusions}
We presented a force-constrained, FT strategy to upscale a DFT-trained MLIP toward QMC accuracy and applied it to simulation of  activated process of mono and multiple S-defects in monolayer MoS\textsubscript{2}. We applied the idea of multi-fidelity learning by using high-level FN-QMC energies and low-level DFT atomic forces to design a FT MLIP. Starting from a robust DFT-trained MACE potential and a sampling pipeline based on perturbed nudged elastic band paths and constrained molecular dynamics, we generated a $\mathcal{O}(10^3)$ set of FN-DMC single-point energies on representative configurations along the minimum energy path. We then fine-tuned the model readout layers with all equivariant message passing layers frozen, minimizing an upscaling loss augmented by a thresholded force-deviation penalty. By fine-tuning only the readout layers of a stable DFT-trained MACE model on FN-DMC energies while constraining deviations from DFT forces via a thresholded penalty, we obtain a practical and controllable approach for QMC-informed potentials in settings where QMC forces are unavailable or too costly.

This construction enables a controlled tradeoff between learning QMC energy corrections and preserving the qualitative DFT force field.
The threshold parameter \(t\) in Eq.~\eqref{eq:forceerror} enables a controllable interpolation between two regimes:
(i) force-anchored: small \(t\) with strongly restricted force changes and potential close to DFT, and 
(ii) energy-corrected: large \(t\) permitting larger corrections and risk of force pathologies.
Because FN-DMC energies are noisy and FN-DMC forces are not used, this parameter is essential to prevent overfitting.
A practical tuning strategy is to increase \(t\) until held-out QMC energy errors stop improving or MD stability degrades.
This tuning can be complemented by freezing most of the network and updating only readouts.
The latter is also conceptually aligned with a ``QMC correction'' to a DFT-learned representation.
Direct training of MLIPs on QMC forces can, when feasible, yield superior force fidelity, as shown in molecular benchmarks.\cite{Slootman2024EthanolQMCforces}
However, for extended systems and/or activated processes in materials where QMC forces are impractical, the present method offers a path to QMC-quality energetics with controlled force behavior.

While the computationally most expensive methods especially in combination with absence of atomic forces, such as QMC, can benefit most from our approach, in principle it can also be used at a lower level of theory, such as DFT. In such a situation one could use first the MACE-Foundation MLIP model~\cite{MACE-Found}, and coarse-grained resample it at the DFT level to construct a higher precision DFT MLIP or just to increase the configuration space coverage.

We have applied the FT MLIP to calculation of energy and free energy migration barriers of mono- and multiple-defects S-vacancies in monolayer MoS\textsubscript{2}. We find that the method achieves a near-FN-DMC accuracy for energies with S-vacancy migration barrier in agreement within $\sim 100\,\mathrm{meV}$ and an estimated MAE atomic force improvement from $220$ for DFT-MLIP to $160$ meV/\AA~for the FT-MLIP. 
These tests have revealed that the FN-QMC potential energy surface is significantly more corrugated compared to DFT, with differences in migration energy barriers of up to 0.55 eV. Contrary, the DFT atomic forces exhibit only minor differences from the FN-QMC benchmark and can be tuned toward FN-QMC accuracy by appropriate scaling, thus confirming the multi-fidelity learning using high-level FN-DMC energies and low-level DFT forces for fine tuning the MLIP. Surprisingly, already a fairly small dataset of FN-DMC energies provided for a significantly improved accuracy of the FT-MLIP model. The entropic contributions were found an important and of the same order of magnitude to the FN-DMC corrections, which in some cases may affect the free energy profiles also qualitatively by locating transition states different from those found from the total energy profiles. We are convinced that the method is completely general as long as inclusion of electronic correlation does not affect the PES qualitatively and is opening the window to calculation of materials with near FN-DMC accuracy with a bearable computational cost. 

\section*{Acknowledgments}
This work was supported by APVV-21-0272, VEGA 2/0133/25, and VEGA 2/0131/23 projects and cofunded by the EU NextGenerationEU through
the Recovery and Resilience Plan for Slovakia under the project Nos. 09I02-03-V01-00012 and 09I05-03-V02-00055. Y. L. was supported by the U.S. Department of Energy, Office of Science, Basic Energy Sciences, Materials Sciences and Engineering Division, as part of the Computational Materials Sciences Program and Center for Predictive Simulation of Functional Materials. We acknowledge the
EuroHPC Joint Undertaking for awarding this project access to the EuroHPC supercomputer LEONARDO through EuroHPC Extreme Access call grants EHPC-EXT-2024E01076 and  EHPC-EXT-2025E01-047. Part of the research results used resources of the National Energy Research Scientific Computing Center, which is supported by the Office of Science of the U.S. Department of Energy under Contract No. DE-AC02-05CH11231, resources at the Center for Nanophase Materials Sciences (CNMS), which is a US Department of Energy, Office of Science User Facility at Oak Ridge National Laboratory under the project No. CNMS2024-B-02618 and resources of the Devana supercomputer project 311070AKF2 funded by European Regional Development Fund under project p37623-1. This work was also supported by the Czech Ministry of Education, Youth and Sports from the Large Infrastructures for Research, Experimental Development and Innovations project IT4Innovations National Supercomputing Center-LM2015070.

\bibliographystyle{unsrt}
\bibliography{references}

\clearpage
\appendix

\setcounter{figure}{0}
\renewcommand{\thefigure}{S\arabic{figure}}
\renewcommand{\theHfigure}{suppfigure.\arabic{figure}}

\setcounter{table}{0}
\renewcommand{\thetable}{S\arabic{table}}
\renewcommand{\theHtable}{supptable.\arabic{table}}

\def\figurename{Supplementary Figure}
\def\tablename{Supplementary Table}
\section*{Supplementary Information}
\section{Details of QMC calculations}\label{app:qmc-details}

The atomic structures and the finite-temperature DFT training datasets for the S-defect(s) in monolayer MoS$_{2}$ were all prepared using periodic setups with 4$\times$4 supercells, see Supplementary Fig.~\ref{fig:supercells}, at the generalized gradient approximation (DFT-PBE)~\cite{pbe} level. 

\begin{figure*}
\centering
\includegraphics[width=1.5\columnwidth,clip,angle=0]{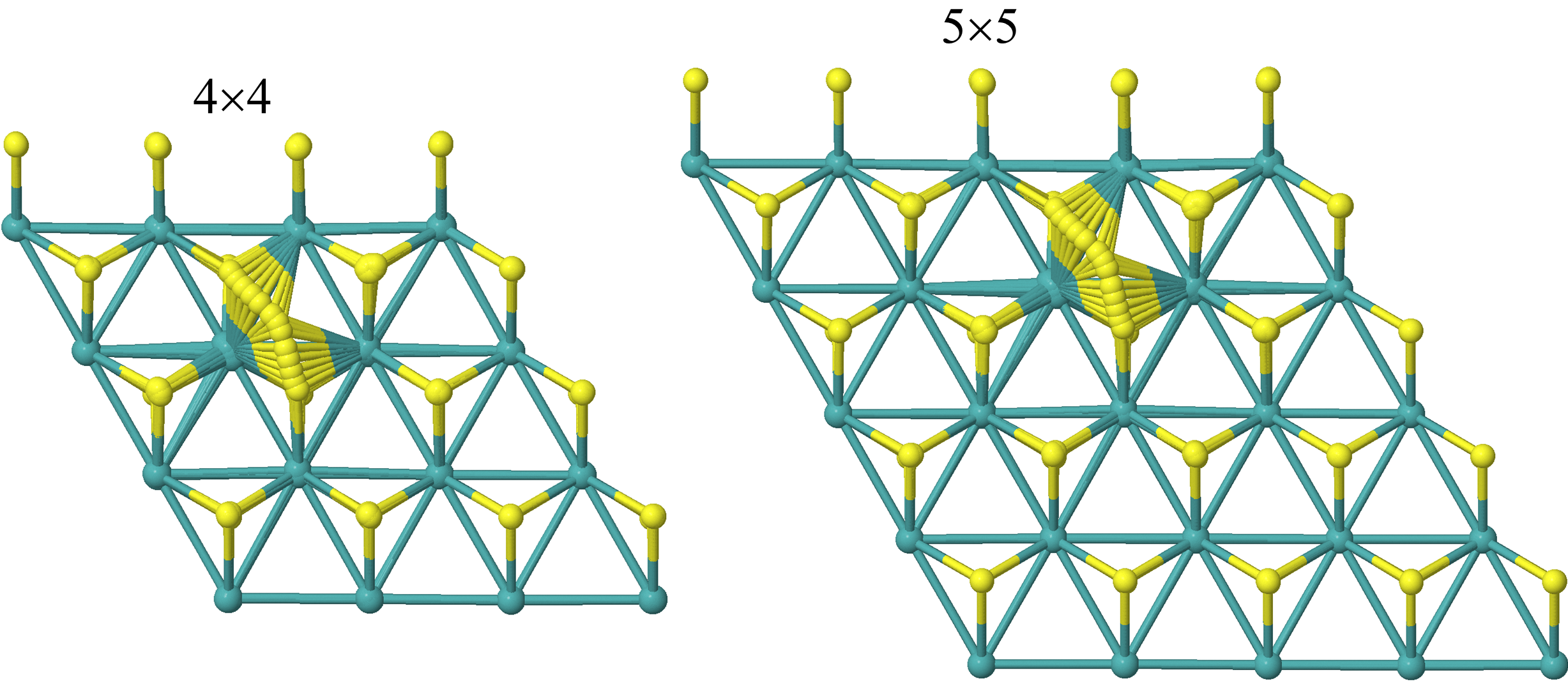}
\caption{
Atomic structure of the S-vacancy in monolayer MoS$_{2}$ enclosed in 4$\times$4 (left) and 5$\times$5 (right) supercells. The Nudged Elastic Band (NEB) minimum energy path for S-vacancy migration between neighboring sites obtained with the fine-tuned MLIP (FT-MLIP), see the main text, is also shown for reference in both supercells.
}
\label{fig:supercells}
\end{figure*}

\begin{figure*}
\centering
\includegraphics[width=1.\columnwidth,clip,angle=0]{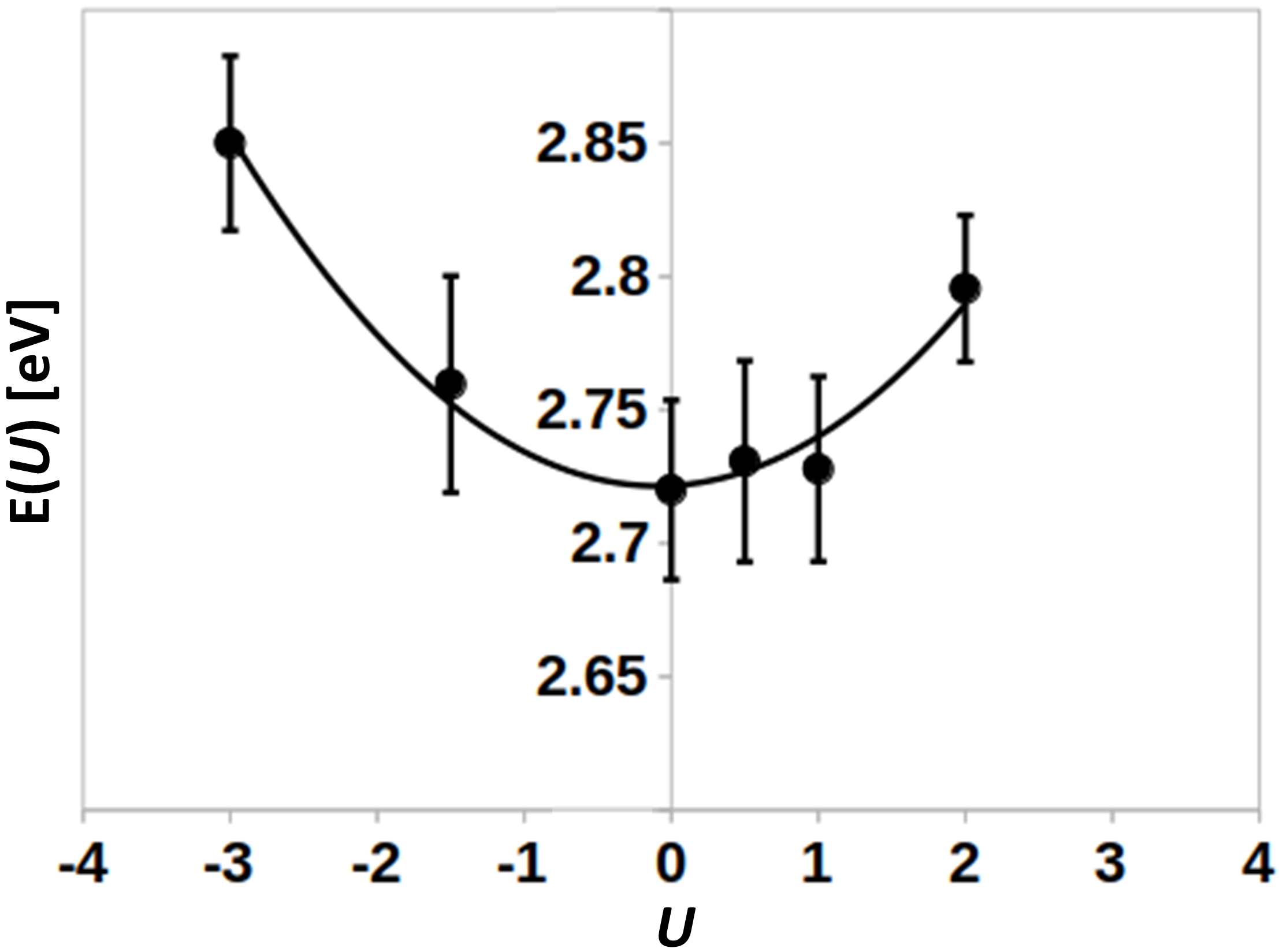}
\caption{
Variation of S-vacancy energy near the transition state as a function of Hubbard $U$. 
}
\label{fig:Hubbard_U}
\end{figure*}

The finite-temperature DFT trajectories were coarse-grained re-sampled at the diffusion Monte Carlo (DMC)~\cite{Foulkes2001QMC} level in fixed-node approximation using variational Monte Carlo (VMC) trial wave functions with the nodal hypersurfaces determined by DFT orbitals using the DFT-PBE electronic orbitals~\cite{pbe}, expanded at the $\Gamma$ point, with short-range correlations described by the Jastrow factor with one- and two-body terms included~\cite{Foulkes2001QMC}. Since Mo contains $d$-electrons, the DFT $d$-electron wavefunctions may be subject to localization errors, we have tested the effect of localization of $d$-electrons via Hubbard $U$. The results are shown in Supplementary Fig.~\ref{fig:Hubbard_U}. The calculated energies indicate that the optimal VMC function corresponds to $U$ = 0. The atomic cores were replaced by Effective Core Potentials.~\cite{mitas_18,mitas_22,ECPs} The effect of the locality approximation was mitigated using the algorithm described in Ref. [\citenumns{casula_06}]. Study of the effect of time-step error on the energetics was studied in Supplementary Fig.~\ref{fig:time_step}. Generally, no to very weak dependence on the time-step length was found and the time-step was set to 0.03 a.u.

\begin{figure*}
\centering
\includegraphics[width=1.5\columnwidth,clip,angle=0]{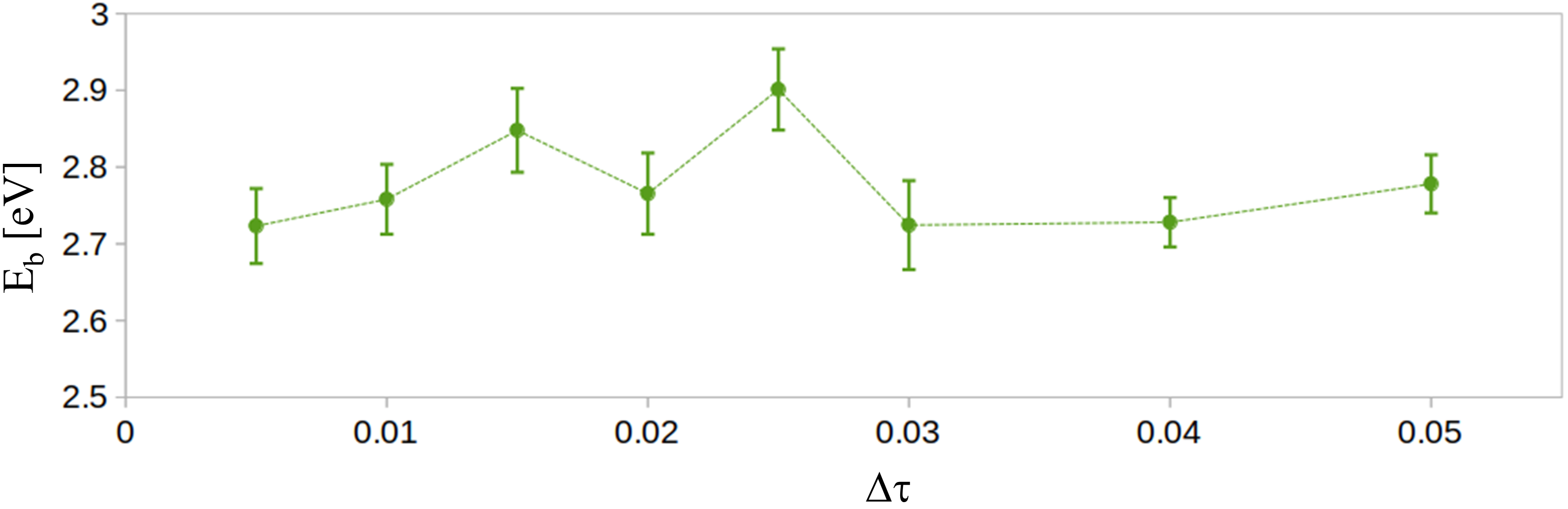}
\caption{
Variation of S-vacancy migration barrier in monolayer MoS$_{2}$ with time step length $\Delta \tau$. 
}
\label{fig:time_step}
\end{figure*}

Since the finite-size bias of FN-QMC energies may be very different from the DFT, we have tested the robustness of our results against the supercell size by calculating the QMC energy differences for a geometry close to the S-vacancy  transition state obtained in 4$\times$4 and 5$\times$5 supercells, see Supplementary Fig.~\ref{fig:supercells}. We found only a negligible difference between the two results: E$_{B}^{\textrm{4}\times\textrm{4}}$ = 2.71$\pm$0.05 vs. E$_{B}^{\textrm{5}\times\textrm{5}}$ = 2.68$\pm$0.08. Hence, all the production calculations were done in the 4$\times$4 supercell and considered well converged w.r.t. the supercell size. 

QMC calculations were performed with the \texttt{QMCPACK} suite of codes~\cite{qmcpack}. All DFT calculations were done with the \texttt{Quantum Espresso} package~\cite{q-espresso}.

To generate the QMC dataset at a manageable cost, we computed FN-DMC single-point energies for $\mathcal{O}(10^3)$ configurations  sampled from the constrained, finite-temperature DFT-PBE MD generated in the main text, i.e. for near-minima and near-saddle point configurations for single S-vacancy migration, with 50\% configurations sampled at 400 K and 50\% at 800 K. The configurations are not uniformly optimized at the FN-DMC level. Rather, we intentionally accept moderately converged DMC energies to maximize configuration-space coverage under a fixed computational budget.
Each production segment consisted of 10 blocks, each block containing 10 DMC steps, with a large 1,008,000 walker population. For most configurations, the run was executed as a sequential restart (36 segments, each containing 25 QMC points/samples). For the initial (0th) configuration, an extended equilibration run of 30 blocks was performed; only the final 10-block segment was retained for analysis. In post-processing, early blocks/segments were discarded to reduce the equilibration bias. For force-related finite-difference analysis, energies were accumulated over 30-block runs to improve the derivatives stability. 

\section{Benchmarks of atomic forces}\label{app:force-benchmarks}

Here we give more details on the more extended atomic force benchmark which complements that shown in Fig. 2 of the main text. For three atoms critical for S-vacancy migration~\cite{Hlozny2025MoS2Vacancies} in configurations samples from a finite-temperature MD corresponding to near minimum, near inflection point, near saddle point, and 3 Cartesian coordinates, 27 displacement profiles in total were computed and fit by quadratic functions as described in the main text. The QMC-derived force components are compared to the corresponding components predicted by the baseline DFT-MLIP and the FT-MLIP models. The results are shown in Supplementary Fig.~\ref{fig:force_benchmarks}. and Tab.~\ref{tab:force_comp}. One can see that at all displacements the DFT energies are significantly displaced from the FN-QMC energies, especially for configurations close to the transition state. Such energy differences account for the different corrugations of the respective PESs. By contrast, the FN-DMC curves and the FT-MLIP model exhibit a high degree of energy agreement. The detailed comparison of all computed energy derivatives, Fig.~\ref{fig:force_benchmarks}, Tab.~\ref{tab:force_comp}, shows that the FT-MLIP corrects especially the large force components (\# 1 15, 17, 21, 23, 26, 27), while worsening only a few small components (\# 6, 19) slightly with respect to DFT-MLIP. Over 27 extracted force components, the baseline DFT MLIP yields a mean absolute error (MAE) of \(220\) meV/\AA\ relative to the QMC derivatives, while the FT-MLIP model reduces the MAE to \(160\) meV/\AA.~
These results indicate that, even though the model is not trained on QMC forces, the energy-driven correction combined with force anchoring can provide a measurable improvement in force fidelity as well.

\begin{figure*}
\centering
\includegraphics[width=1.8\columnwidth,clip,angle=0]{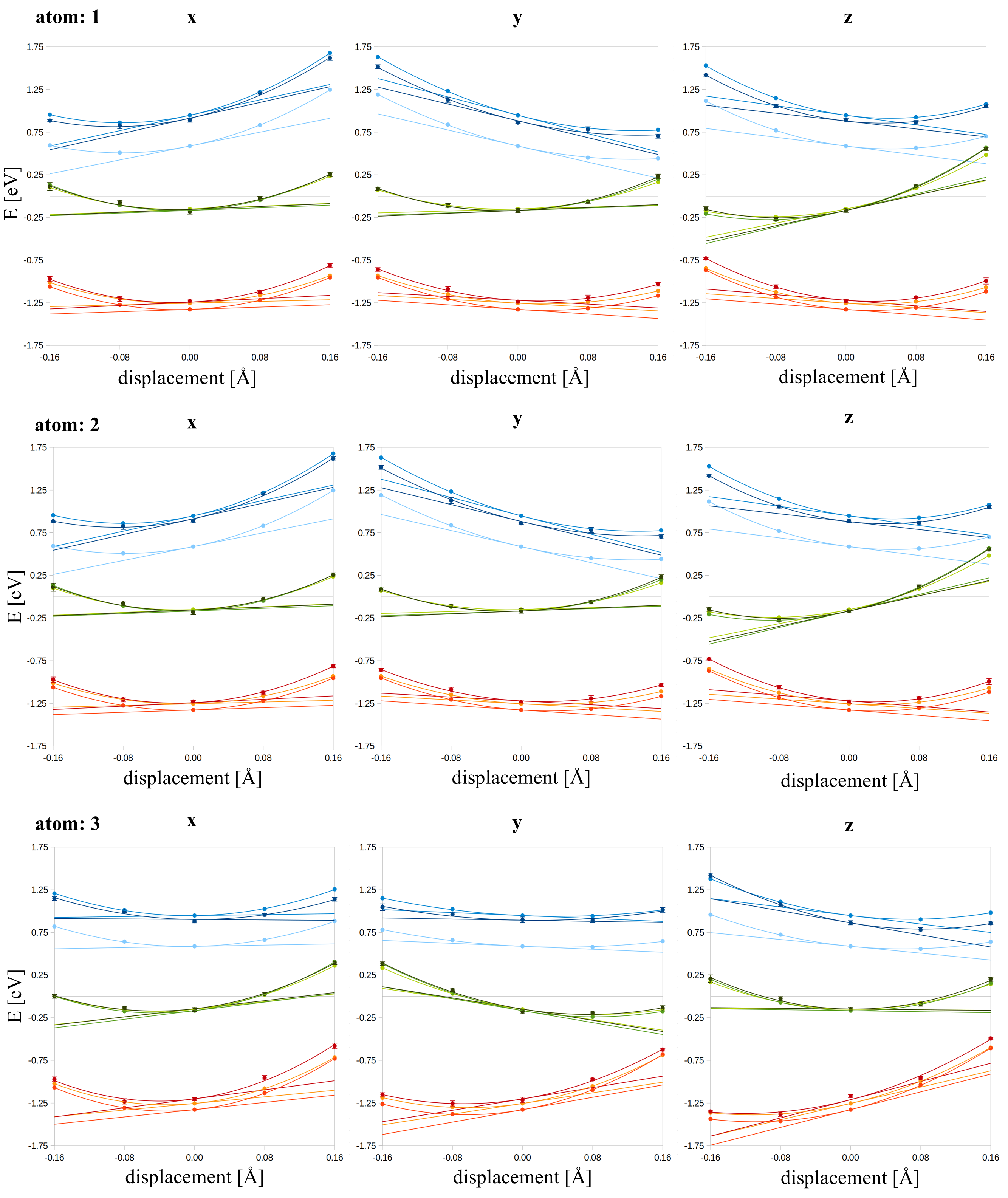}
\caption{\textbf{Extensive force benchmark against FN-DMC energy derivatives.} Example of the finite-difference benchmark of FN-DMC force components calculated for three different atoms, x, y, z components and configurations from a finite-temperature dynamics near the energy minimum (red curves), near inflection point of the corresponding NEB trajectory (green curves), and near the transition state (blue curves). For a fixed configuration, a target atom was displaced along one coordinate in five steps (\(\Delta=0.08\)\AA) and FN-DMC energies were computed for each displacement (dark color curves). Quadratic fits yield the derivative at the origin, providing a FN-DMC estimate of the force component. The baseline DFT-MLIP (lighter colors) and the FT-MLIP (light colors) are evaluated on the same displaced structures to compare their implied derivatives. Note that all energies are aligned to an arbitrary configuration near the inflection point of the corresponding NEB.
}
\label{fig:force_benchmarks}
\end{figure*}

\begin{table}[h!]
\centering
\caption{\textbf{Comparison of atomic forces.} Summary of all atomic force components extracted as outlined in Fig.~\ref{fig:force_benchmarks} by FN-QMC, FT-MLIP, and DFT-MLIP.
}
\label{tab:force_comp}
\begin{tabular}{l|c|c|c|c|c|c|c|c|c|c}
\hline
\hline
method & \multicolumn{10}{c}{force components [eV/\AA]} \\
\hline
configuration & \multirow{2}{*}{comp.} &  \multicolumn{3}{c|}{1} & \multicolumn{3}{c|}{2} & \multicolumn{3}{c}{3} \\ \cline{1-1} \cline{3-11}
atom &   & 1 & 2 & 3 & 1 & 2 & 3 & 1 & 2 & 3 \\
\hline
\hline
FN-DMC   & x &  2.31 & -0.07 & 0.46 & 0.50 & 1.32 & -0.88 & 0.42 & 1.19 & 1.02 \\
FT-MLIP  & x &  2.25 &  0.14 & 0.18 & 0.34 & 1.06 & -1.08 & 0.39 & 1.26 & 0.75 \\
DFT-MLIP & x &  2.03 &  0.18 & 0.02 & 0.25 & 0.97 & -1.15 & 0.41 & 1.12 & 0.47 \\
\hline
FN-DMC   & y & -2.47 & -0.17 &  0.01 & -0.56 & 1.67 & -3.03 & 0.43 & -1.64 & 1.60 \\
FT-MLIP  & y & -2.68 & -0.43 & -0.05 & -0.67 & 1.81 & -2.97 & 0.35 & -1.74 & 1.68 \\
DFT-MLIP & y & -2.35 & -0.43 & -0.07 & -0.56 & 1.56 & -2.60 & 0.29 & -1.53 & 1.42 \\
\hline
FN-DMC   & z & -1.15 & -1.77 & -0.26 & -0.82 & 2.67 & 0.31 & 2.24 & -0.10 & -1.58 \\
FT-MLIP  & z & -1.41 & -1.24 & -0.56 & -0.78 & 2.60 & 0.24 & 2.43 & -0.14 & -1.50 \\
DFT-MLIP & z & -1.29 & -1.00 & -0.55 & -0.69 & 2.39 & 0.19 & 2.07 & -0.07 & -1.30 \\
\hline
\hline
\end{tabular}
\end{table}

\end{document}